# Understanding Data Search as a Socio-technical Practice

Kathleen M. Gregory[1], Helena Cousijn[2,3], Paul Groth[4,5], Andrea Scharnhorst[1], Sally Wyatt[6].

[1]Data Archiving and Networked Services, Royal Netherlands Academy of Arts and Sciences, The Hague, The Netherlands; [2]Elsevier, Amsterdam, The Netherlands, [3]DataCite, Hannover, Germany; [4]Elsevier Labs, Amsterdam, The Netherlands ; [5]University of Amsterdam, The Netherlands; [6]Maastricht University, The Netherlands

## Abstract

Open research data are heralded as having the potential to increase effectiveness, productivity, and reproducibility in science, but little is known about the actual practices involved in data search. The socio-technical problem of locating data for reuse is often reduced to the technological dimension of designing data search systems. We combine a bibliometric study of the current academic discourse around data search with interviews with data seekers. In this article, we explore how adopting a contextual, socio-technical perspective can help to understand user practices and behavior and ultimately help to improve the design of data discovery systems.

## Introduction

The reuse of open research data is heralded as having the potential to increase effectiveness, productivity, and reproducibility in science.[1,2] However, data do not flow easily between users, situations, and disciplines.[3] Instead, they rely on dynamic relationships between people, context and technology.

We aim to explore these relationships in an ongoing project that integrates science and technology studies and information science to inform and intervene into the design of tools for searching for research data. In this paper, we answer recent calls to integrate scientometric studies and qualitative methods [4] by combining bibliometrics and interviews to gain empirical evidence about researchers' data search practices.

Interest in facilitating data sharing and reuse is high. Funding agencies, research organizations, and repositories are all increasingly engaged in drafting policies regulating data sharing and management.[5] Many studies mirror recent policy developments, focusing on data sharing and management practices, usually presenting the researcher in the role of data creator.[6,7] As more work is done to investigate how data are used, researchers' multiple roles are being recognized. Researchers are not only data producers, but they also act as consumers, curators, and evaluators of data.[8]

Before data can be reused, they must first be discovered. Researchers seek, access, and evaluate data they have not created themselves as they engage in the process of searching for data. Data search has recently emerged as a separate topic of inquiry within the core information retrieval community.[9] Here, research has focused primarily on finding technical solutions,[10] with the development of ontologies, standards, and search tools taking precedence.[11,12] Investigations into the social aspects of data search, such as data discovery and reuse practices, are far less common.[13,14]

Our aim is to build on these investigations to understand data search from a socio-technical perspective. How do data (re)users locate and make sense of research data, and for which purposes? How are these practices situated with regard to technical resources and within communities?

We tackle these questions using two methods. We begin by reviewing the data search literature with a bibliometric analysis, further revealing the technical bias, the distributed nature of the discourse, and gaps in terms of data search processes. We then present the results of interviews across disciplinary domains. The interview questions, informed by both our bibliometric study and dialogue with the designers of a data search system, explore data users' needs and contexts, their strategies for locating data, and the criteria brought to bear when evaluating reuse potential. Inspired by the quantitative analysis, we present our qualitative interview data using a unique tabular presentation

In our analysis, we view context as a collection of interacting, dynamic components,[15] where users are often simultaneously embedded in multiple contexts with varying social norms.[16] We also draw on the conceptualization of users as "social actors",[17] who internalize and act on the social and information norms of their communities (as summarized by Courtright).[18] We conclude by discussing how data search can be understood as a socio-technical process rooted in context and pose suggestions for integrating these insights into the design of data discovery systems.

## Bibliometric Study and Analysis – Method I

Research on data search currently focuses on technical challenges and solutions for searching data, as techniques for document-based retrieval do not work well for structured data.[19] In response, researchers seek ways to apply keyword searching to datasets,[20] to semantically enhance datasets,[21] and to create new ontologies and standards.[11,12] These approaches are used to create a variety of search tools for specific disciplines[22,23] and data types.[24,25]

Studies of users' practices are not as prevalent. A small body of work examines how users seek and evaluate data within disciplines,[26,27,28] across data-related professions[29] or within data repositories.[30,8,31,32] Much information on data search practices is buried within investigations of other data behaviors, such as studies investigating the characteristics of data sharing and (re)use in specific research teams and disciplines.[33,34,35,36] Work investigating the qualities of successful data reuse[37] and that examine criteria that researchers use to establish data trustworthiness (e.g. the identity of the data creator, the reputation of a repository, or prior usage)[38] is especially relevant when examining how users evaluate and make sense of data. Trust development in particular is recognized as a complex, non-linear, social enterprise.[39]

A broader bibliometric analysis of the data search discourse makes this imbalance between technical and social research explicit. Building upon our earlier review of observational data users,[10] we searched the literature using different keyword combinations across all fields, primarily in Scopus. We combined keywords related to information retrieval (e.g. user behavior, information seeking, information retrieval), data practices (e.g. research practices, community practices) and research data (see Figure 1). We performed other keyword searches for data search and discoverability and applied bibliometric techniques such as citation chaining and related records. Pertinent sources (journals, book series, proceedings, etc.) not indexed by Scopus were searched directly using similar keywords. We closely read the 400 retrieved documents to identify relevant publications, resulting in a final corpus of 189 documents published between 1990 and 2017.

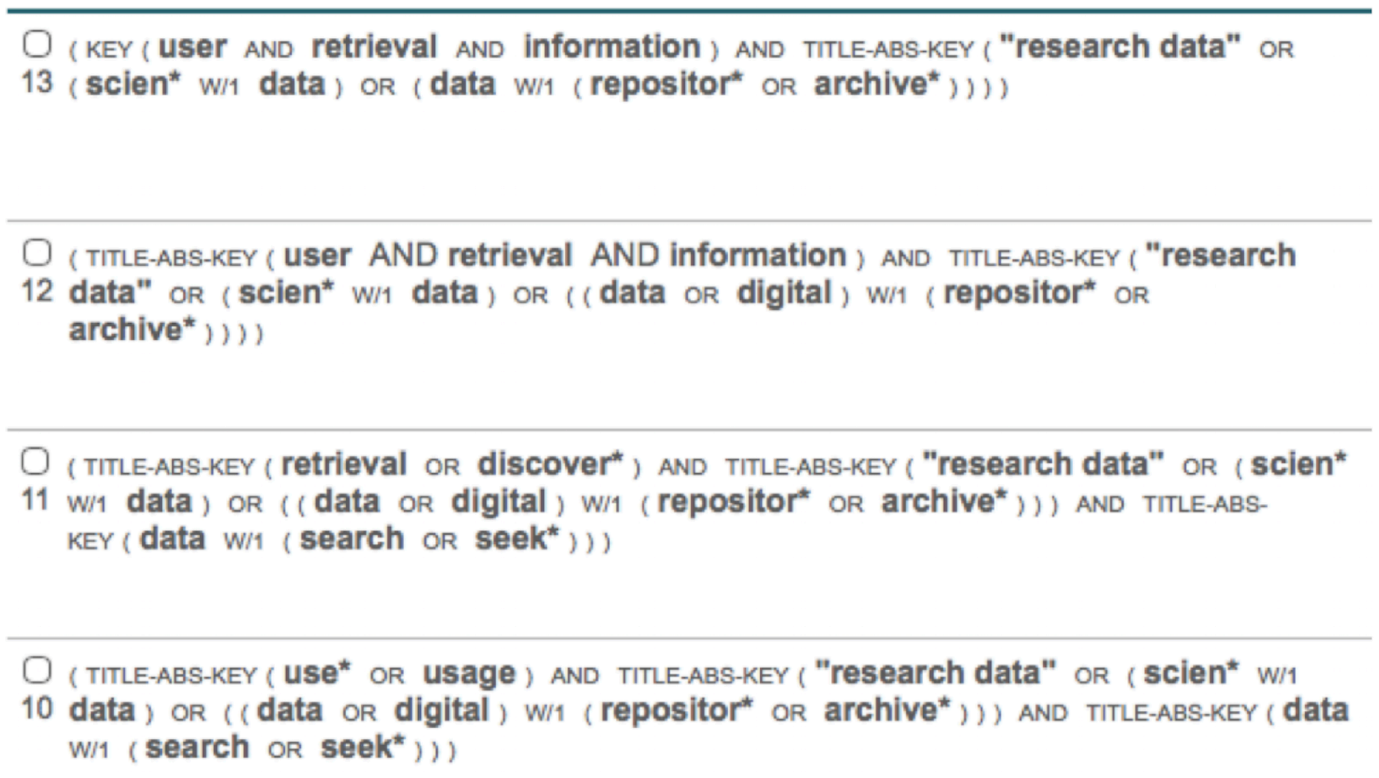

*Figure 1. A sampling of keyword combinations used in Scopus*

Restricting our results to journals and conference proceedings produced a corpus containing 102 sources titles and 182 publications. As Bradford's law predicts,[40] a small, core group of sources accounts for the majority of publications; only ten sources have published more than three publications. Most publications in these ten sources were published after 2012, indicating that interest in the topic has only recently begun to solidify (Figure 2). Scopus classifies seven of these top sources as computer science or engineering titles, highlighting the technical dominance in the discourse. We also see some hint of discipline-specific interest, evidenced by the appearance of *Nucleic Acids Research* and *Bioinformatics*.

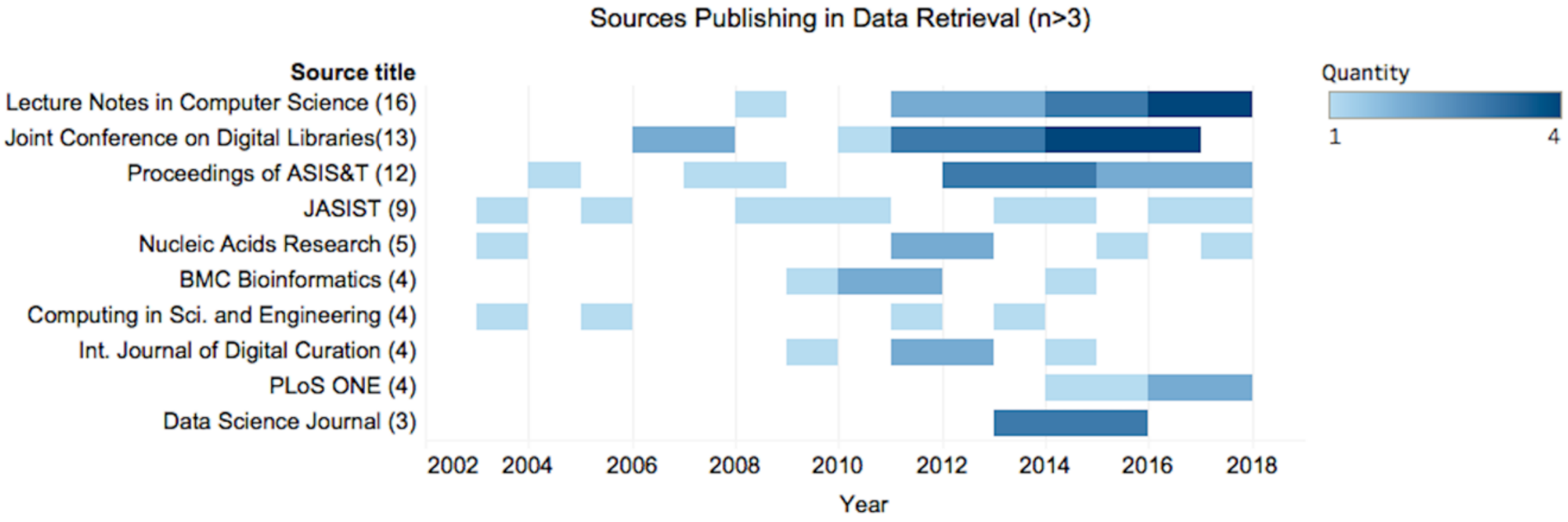

*Figure 2. Sources in the corpus with more than three publications*

The publications in Figure 2 include approximately 40% of the corpus. The remaining 60% are spread across other sources, including disciplinary journals which only appear once. The two earliest publications come from molecular biology and astronomy, indicating an early interest within traditionally data-intensive disciplines.

To augment the source-level analysis, we created a keyword co-occurrence map for all keywords using VOSviewer[41] (Figure 3). The minimum co-occurrence was set to five; general keywords such as "article" or "procedures" were excluded.

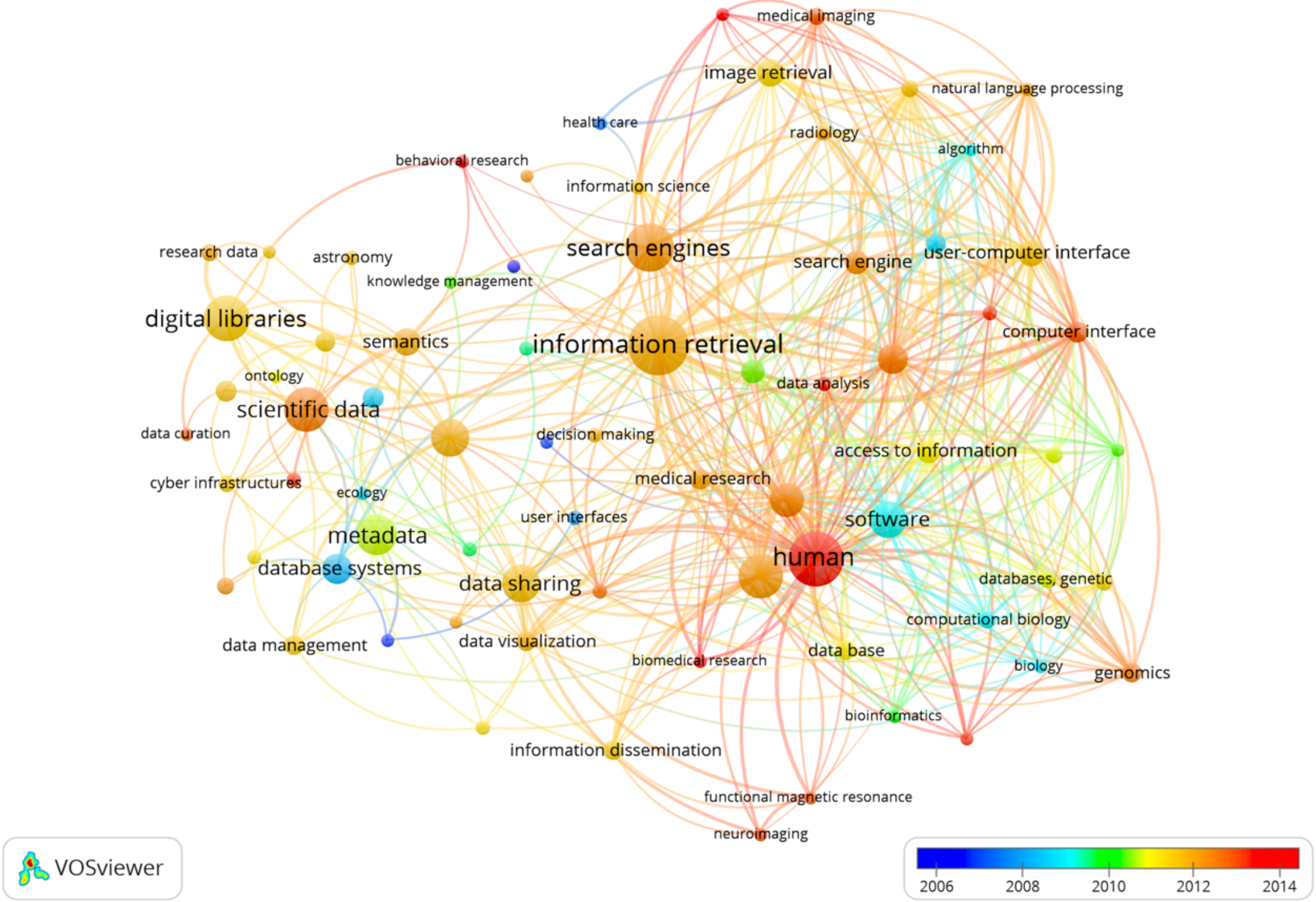

*Figure 3. Keyword co-occurrence map with temporal overlay*

Figure 3 demonstrates the spread of the discourse and its technical focus. We see the dominance of disciplinary keywords (e.g. genomics, astronomy) and the prominence of more technical areas (information retrieval, search engines, and image retrieval). The concept of human solidifies only recently; the highest average number of keywords appears in 2014, both in connection to technical concepts (information retrieval, search engines) and keywords associated with the study of humans (neuroscience, genomics).

The keyword mapping highlights gaps in the field. Only a handful of disciplinary keywords are present, mirroring the disciplinary bias in the source analysis. Fields with well-developed data infrastructures, such as high-energy physics,[42] are surprisingly under-represented as are the social sciences. “Data search” or “data retrieval” are not identified as stand-alone topics, suggesting that information about these practices is buried within other discussions and that these terms are not yet codified labels. The analysis also suggests that social factors have only recently begun to be addressed.

Kacprzak and colleagues echo this idea, noting that existing data search applications are often based on preliminary or anecdotal evidence.[43] Query log analyses can help to develop insights about data search practices,[43] but log data cannot explain the reasoning behind search behaviors.[44] Our work begins from a deeper engagement with users themselves, conducting interviews informed by the findings of this bibliometric analysis and conversation with creators of a data search system. We thereby attempt to facilitate a dialogue between system designers and end users, bridging the social and the technical aspects of the problem while using scientometrics to inform qualitative research.

## Interview Design and Analysis – Method II

### *Development of Interview Protocol*

We drafted our interview questions to draw out the data needs, search practices, and evaluation behaviors of participants in a variety of disciplines. The development of our interview protocol was informed by the bibliometric study, our previous survey of the data search literature[10] and consultations with the Elsevier Data Search (datasearch.elsevier.com) design and implementation team. Data Search is a publicly available data search engine that harvests data from multiple data and publication repositories.[45] It is currently under development; its design is not yet finalized. In the development process, the design team has conducted preliminary evaluations of user needs and behaviors. We used these evaluations as input to our interview protocol.

### *Interviews*

We conducted 22 one-hour, semi-structured interviews between October and December 2017, using Skype, GoToMeeting or in-person meetings. Participants were recruited via email from a pool of 186 individuals who had visited the Data Search portal and had indicated willingness to provide feedback. We postulated that these individuals would have interest in searching for research data. We spoke with 19 respondents from the Data Search pool and recruited three additional participants using convenience sampling. We obtained ethical review for the study, and all participants provided written informed consent.

The majority of participants are active researchers, although some are active in other areas or have numerous roles (Table 1); participants also spoke about previous experiences in other roles or the experiences of their colleagues. Participants work in twelve countries. The most frequently represented countries are the United States (n=6) and the Netherlands (n=3). Some participants currently work outside of their home countries or have past experience working abroad.

Participants self-identified their broad disciplinary area. Although a range of disciplines are present, computer science (n=3) and information science (n=3) are the most common. Individuals working in support roles, especially in libraries, have insight into the needs and practices of various disciplines (n=3). Our participants are in diverse career stages: early career (0-5 years, n=5), mid-career (6-15 years, n=10), experienced (16+ years, n=6), and retired (n=1).

| Participant | Gender | Discipline | Country of employment | Career Stage | Current role |
|---|---|---|---|---|---|
| 1 | M | Medical statistics | USA | Retired | Citizen |
| 2 | M | Business and finance | UK | Experienced | Industry |
| 3 | F | Cognitive psychology, neuroscience | Netherlands | Early | Researcher |
| 4 | F | Evolutionary ecology | France | Early | Researcher |
| 5 | M | Information science | Singapore | Early | Researcher |
| 6 | F | Water resources | Malaysia | Early | Researcher, PhD candidate |
| 7 | F | Computer science | Australia | Middle | Researcher |
| 8 | M | Computer science | USA | Middle | Researcher |
| 9 | M | Information science | Spain | Middle | Researcher |
| 10 | M | Information science, musicology | Netherlands | Middle | Researcher |
| 11 | M | Psychology | Spain | Middle | Researcher |
| 12 | M | Psychiatry, medicine, neuroscience | Portugal | Middle | Researcher, clinical practice, PhD candidate |
| 13 | F | Cellular/molecular biology, medical devices | USA | Middle | Researcher, industry |
| 14 | M | Computer science, data science | Guatemala | Middle | Researcher, industry |
| 15 | M | Acoustical engineering | Canada | Experienced | Researcher |
| 16 | M | Industrial ecology | Brazil | Experienced | Researcher |
| 17 | M | Paleontology | Netherlands | Experienced | Researcher |
| 18 | F | Popular culture | USA | Experienced | Researcher |
| 19 | M | Chemistry | Australia | Experienced | Researcher, industry |
| 20 | F | Libraries | USA | Early | Support |
| 21 | M | Libraries | UK | Middle | Support |
| 22 | F | Scientific literature manager | USA | Middle | Support |

*Table 1. Participant description*

Although participants were recruited from Data Search, the interviews did not focus on this tool. Rather, our aim was to learn more about general data search practices. We therefore asked open-ended questions, encouraging rich discussions about contexts and data needs, strategies for locating data and criteria for evaluation. We focused on data not created by participants, but otherwise left the term “data” open for definition by participants.

When necessary, we followed our questions by prompts to elicit more detail, some reflecting the interests of the search engine team. We audio-recorded the interviews and created detailed summaries for each interview. These summaries were uploaded into the qualitative data analysis program QDA Miner Lite for coding and analysis.

## Interview Findings

Our main findings are presented along three dimensions, informed by the analytical framework we have discussed elsewhere.[10] These three dimensions - user contexts and data needs, search strategies, and evaluation criteria - are further divided into non-hierarchical subsections designed to orient the reader to the interpretations presented in the discussion. We summarize our interview findings in tabular form; we do this not to indicate statistical meaning, as our sample size is small. Rather, we use the tables to provide an overview that facilitates navigating the discursive analysis of our interview findings.

### *User Contexts and Data Needs*

An important contextual aspect is a user's disciplinary community. For some, data sharing and reuse are normal, as in computer science, biomedicine, and astronomy (7, 13, 20). For others, seeking data involves calculated exchanges, where information is only shared with individuals who share reciprocally (2).

Within disciplines, ideas of data ownership, data sharing regulations, data types, and cultural differences also affect participants' practices.

> We have to be very careful sharing the raw data, otherwise we can be sued by the lab or by the university...In terms of the clinical data (in Portugal), the situation is different…it is not that formal as in the United States. It is quite frequent to share clinical data with other researchers (12).

A lack of standardization in collection, description and sharing practices, even within disciplines where data sharing is established, can negatively affect data discoverability, evaluation, and reuse (8).

The data user is not always the person searching for data. Experienced researchers delegate some search responsibilities to graduate students, training students to find data for background purposes primarily by searching the literature (13, 16). Librarians and those in support roles assist students (20), external researchers (21), and those working in industry (22) to locate data.

Not all data seekers are involved in research. One participant does not have an academic research background but is interested in finding data from medical studies in the press (1). Others work in industry; an increasing number of students are seeking data (20).

#### *Diverse and changing needs*

Participants need a variety of data for a variety of purposes. Chemists need numerical data such as superconducting temperatures and non-numerical spectral data; an evolutionary ecologist requires physiological data and field observations; and a humanist seeks social media posts for textual analysis. Data are needed for background purposes supporting research (Table 2) and for foreground purposes (Table 3) driving new research.

| Discipline | Comparison | Evaluate systems / algorithms | Overview of new topic / method | Learning data-centric skills | Supporting others | Verification of own results | Make predictions | Teaching/ student learning | Proposal preparation |
|---|---|---|---|---|---|---|---|---|---|
| Acoustical engineering | 15 | | | | | 15 | | | |
| Business & finance | | | 2 | | | | | | |
| Ecology - industrial & evolutionary | | | 16 | | | | | 4 | 4 |
| Information science / Computer Science | 9, 10 | 5, 7, 8, 14 | | 9 | | 8 | | | |
| Librarians/ Literature managers | | | | | 20, 21, 22 | | | | |
| Molecular biology | 13 | | | | | | 13 | | |
| Psychiatry, medicine, neuroscience | 12 | | 12 | | | | | | |
| Psychology, Cognitive psychology, neuroscience | | | | 11 | | | | | |
| Water resources | | | | | | 6 | | | |
| Astronomy (reported) | 20 | | | | | | | | |
| Students (reported) | | | | | | | | 20 | |

Researcher Support Industry Student

*Table 2. Background purposes for seeking data*

| Discipline | Model/ system inputs | New questions of existing data | Research using social media | Identify new substance, organism | Replication |
|---|---|---|---|---|---|
| Chemistry | | | | 19 | |
| Ecology - industrial & evolutionary | 4, 16 | 4, 16 | | | |
| Information science / Computer Science | 7, 8, 14 | 9 | 2, 9 | | |
| Medical statistics | | | | | 1 |
| Paleontology | | | | 17 | |
| Popular culture | | | 18 | | |
| Psychology, Cognitive psychology, neuroscience | | 11 (planned) | | | 11 |
| Water resources | 6 | | | | |
| Climate change (reported) | | 21 | | | |
| Sociologiy (reported) | | 21 | | | |

Researcher Citizen

*Table 3. Foreground purposes for seeking data*

Participants need data from their own disciplines, but they also require other data. Across domains, participants use data created on social media platforms or data produced by governments. Researchers are also interested in data created for another purpose in other disciplines, e.g. sociologists seeking data collected by female freshwater biologists in the early 20$^{th}$ century (21), or engineers building a biomimicry database (17).

Descriptions about data can be as important as the data themselves. Paleontologists rely on descriptions about physical specimens in the literature to trace evolution over time. Specimens are sometimes destroyed or inaccessible, bestowing extra value to these descriptions (17).

Metadata are crucial for computer and information scientists, who need rich metadata to design and test systems.

> I am not really using observation values. I am more using the description of the datasets, the metadata…Also I am using secondary metadata. The datasets are linked to publications, and I am using metadata about the publications (7).

At other times, computer and data scientists need large multidisciplinary datasets with minimal indexing to create indexing systems or to test algorithms (14).

Research interests determine the data that participants need. As interests change, so do data needs. Sometimes these changes result from new subject interests (11) or job changes (4). Needs can also evolve in response to the changing nature of data and the research environment. As more data become openly available, some researchers are curious about using open data or experimenting with data science approaches:

> A new approach would be pulling large datasets into some data mining process. This is something that big data scientists do. Sometimes they find specific relationships between variables, that apparently don't have anything to do (with each other). They don't know how to interpret these relationships. But we psychologists and social scientists actually have the theory and the background to interpret these relationships. (11)

These projects demand datasets of increasing size and require researchers to develop new data analysis skills (9).

Participation in interdisciplinary projects is common. This is mirrored in an observed increase in interdisciplinary research institutes at the university level; these institutes provide an opportunity for researchers across disciplines to encounter and use each other's data (21).

## Search Strategies

### ***Multiplicity of resources and strategies***

Most participants employ a mixture of strategies and visit multiple resources to find data (Tables 4 and 5). The majority use Google either to search for data or to locate repositories; the success of these searches is mixed.

> I just search for keywords (in Google]) - prostate, MRI, segmentation...but it is not very effective. I can't remember a single instance when I found something useful. What people don't realize is that in medical imaging, data has to be in a special format. If you just search for prostate images, you will see a lot of pretty pictures, JPEGS, but that is not suitable for the analysis (8).

| Discipline | People | Search engines | Disciplinary data repositories & collections | Literature | Government data portals | Platforms w/ user-created data | Websites | General data repositories | Museums, libraries | Data vendors | Internal systems | Social media forums | Industry associa-tions |
|---|---|---|---|---|---|---|---|---|---|---|---|---|---|
| Acoustical engineering | | | | 15 | | | | | | | | | |
| Business & finance | 2 | 2 | | 2 | 2 | | | | | | | | 15 |
| Chemistry | 19 | | 19 | | | | | | | | | | |
| Ecology - industrial & evolutionary | 4, 16 | 16 | 16 | 16 | | | | | | | | | |
| Information science / Computer Science | 5, 9, 14, 7, 10 | 5, 14, 8, 10 | 8, 10 | 5 | 9 | 7, 9, 14 | | 14 | | 5 | 8 | 5 | |
| Librarians/ Literature managers | 22 | 22 | | 22 | | | 20 | 20 | | | | | |
| Medical statistics | 1 | 1 | | 1 | | | | | | | | | |
| Molecular biology, medical devices | 13 | 13 | 13 | | | | 13 | | | | | | |
| Paleontology | 17 | | 17 | 17 | | | | | 17 | 17 | | | |
| Popular culture | 18 | 18 | 18 | | | 18 | 18 | | 18 | | | | |
| Psychiatry, medicine, neuroscience | 12 | | | | | | | | | | 12 | | |
| Psychology, Cognitive psychology, neuroscience | 11 | 11 | | | 11 | | | | | | | | |
| Water resources | 6 | 6 | | | 6 | | | | | | | | |
| Astronomy (reported) | 20 | | 20 | | | | | | | | | | |
| Biology (reported) | 21 | | | | | | | | | | | | |
| Historians (reported) | 20 | | | | | | | | | | | | |
| Sociologists (reported) | | | 21 | | | | | | | | | | |
| Students (reported) | | 20 | | | 20 | | | | | | | | |

| | | | | |
|---|---|---|---|---|
| Researchers | Support | Industry | Citizens | Students |

*Table 4. Resources for seeking data*

| | Literature | | | Keywords | | Push | | | Browsing | | | Other | |
|---|---|---|---|---|---|---|---|---|---|---|---|---|---|
| Discipline | Accession number/reference to datasets for follow-up | To find specific technique, parameters | Use figures, graphs, tables, supplementary materials | With data terms/ formats/ known dataset | Subject terms. only | Twitter feed | Newsletter or discussion list | Standing literature search | Government repositories/ cultural site | Small disciplinary or clinical repository | Twitter, social media (trending, key voices) | API | Social interactions - discovery and access |
| Acoustical engineering | | 15 | | | | | | | | | | | |
| Business & finance | | | 2 | | | | | | | | | | 2 |
| Chemistry | | | 19 | | | | | | | | | | 19 |
| Ecology - industrial & evolutionary | 4 | 16 | | | 16 | | | | | | | | 4, 16 |
| Information science / Computer Science | 5, 8, 10 | | | 5, 8, 10 | | 9 | 14 | | 9 | 10 | | 5, 14, 7, 8 | 9, 10, 14 |
| Librarians/ Literature managers | | | | | 20 | | | 22 | 20 | | | | 22 |
| Medical statistics | | | | 1 | | | | 1 | | | | | 1 |
| Molecular biology, medical devices | 13 | | | | | | | | | | | | 13 |
| Paleontology | | | | | | | 17 | | | | | | 22 |
| Popular culture | | | | | 18 | 18 | | | | | 18 | | 18 |
| Psychiatry, medicine, neuroscience | | 12 | | | | | | | | 12 | | | 12 |
| Psychology, Cognitive psychology, neuroscience | | | | 11 | | | | | | | | | 11 |
| Water resources | 6 | | | | | | | | 6 | | | | 6 |
| Astronomy (reported) | | | | | | | | | | | | | 20 |
| Biology (reported) | | | | | | | | | | | | | 21 |
| Historians (reported) | | | | | | | | | | | | | 20 |
| Sociologists (reported) | | | | | | | | | | | | | 21 |

Researchers Support Industry Citizens

*Table 5. Strategies to locate data*

Disciplinary data repositories are important sources. Certain resources are considered to be "gold standards" within disciplines. Researchers also discover resources as they create data management plans or share their own research data in repositories (21,4).

Many individuals discover data serendipitously while reading the literature; fewer actively search for data by conducting literature searches. Those who do are outside academia (1, 22, 2), see the literature as part of their data (5, 17) or only need values commonly reported in the literature (15, 16). When participants encounter data serendipitously in the literature, they follow up by contacting authors or using dataset accession numbers or titles to locate the dataset online.

Other search strategies (Table 5) include keyword searches (with and without data-specific terms) and browsing using metadata or graphical interfaces. Some participants feel limited by keyword-only search boxes and desire greater browsing capabilities to increase precision.

Participants also find data through push-based strategies such as saved literature searches, subscriptions to newsletters or discussion lists, and carefully constructed Twitter feeds. Often these services are set up to meet other information needs, and participants discover data through them serendipitously (9, 14, 17). Other participants create feeds with the explicit purpose of finding textual data or data in the literature (18, 22).

Computer and information scientists use application program interfaces (APIs) to efficiently gather large amounts of data. They envision searching capabilities that incorporate APIs and link with computational tools (8) and systems that proactively search for data and present them in new ways.

***Social interactions in finding data***

Participants locate data from colleagues, collaborators, supervisors, data authors, and support staff both serendipitously and intentionally. Data are encountered serendipitously during conference presentations (19, 13, 22), informal conversations with colleagues (3) or because of geographic proximity (21). Interdisciplinary networking and training events provide an opportunity to unexpectedly learn about others' data; some interactions lead to data reuse.

> After these events, there is a lot of time for networking, and we have seen collaborations starting…between the School of Computing, who do the (data) visualization, and Sociology, who create interesting datasets and are interested in visualizing their data (21).

Personal connections are the most efficient and accurate route to data search for some participants.

> Actually, most of the times that I have looked for external data, it has been through (personal) connections (11).
>
> The human network of contacts is still the best way to find the information you want, especially if it is a small group...that is the most powerful and accurate source of information that I use at this point. (17)

Support staff engage in dialogue with their patrons to more accurately locate needed data (20, 22). Researchers also seek input from colleagues to design efficient queries (14). Some participants educate colleagues about the limitations of databases and how to search effectively (13, 20, 16).

Professional networks and connections are also key to accessing data once found. Access to medical records and images is only possible through hospital affiliations (12, 8). Data are viewed as a gift in some communities; access is only granted to a small trusted circle of colleagues (20, 2). Historians must develop close relationships with the family members of people they are researching in order to be given access to documents (20). In countries with developing digital infrastructures, personal connections can be the only way to access non-digital data.

> In my country, you don't get all of the data online. Sometimes you need to do the personal approach with some people in the (governmental) agency or department…I went directly to the department and met the person in charge, and then it was easier (6).

Collaborations provide a safe way for researchers to share and access data (3, 7). If an industrial ecologist needs data collected by industry or from another discipline, he forms new collaborations (16). Collaborations developed to access data can help early career researchers to grow their professional networks (4, 6).

Finding data through social connections has limitations. Researchers risk operating in "filter bubbles" by only seeking information within their network (3, 18). Some assume there is no valuable data outside of their circle (21). Even within one's own lab, one may not know the details of others' data (8). It can be especially difficult to know who to contact to obtain data when operating outside of one's area of expertise (12, 14).

***Success***

Participants using literature searches to locate data are satisfied with their methods or cannot think of better ways to meet their needs (1, 5, 15, 16, 22). Researchers seeking data in other ways also feel that their methods are sufficient (11, 6, 4, 7), and believe they must combine multiple strategies to be successful (6), especially when they have goals beyond locating data.

> I think if there was a good search engine, then I could get the dataset directly. I would still get in touch with the data author anyway, both for social reasons - developing the network and eventual collaboration - and also because most of the times the metadata are not enough to really understand the biology behind the species (4).

If researchers cannot find the data they seek, they assume that they are not available online (17, 4). Some participants will make do with the data they find or will give up the search (14, 10, 2). Others will create their own data, believing that the data they need do not exist (9, 10).

Searching skills also affect success. Students are often unskilled at finding and evaluating data (20). Although experts are assumed to be highly skilled at finding data in their field, this is not always true (21).

### *Not all data are findable*

Metadata quality determines whether or not data are findable (8, 21). Even in disciplines with well-developed metadata standards, researchers do not always describe their data in ways that facilitate discovery (8) or follow standardized sharing methods (14, 10). When researchers do follow best practices, discovery can be hindered by poor links between data and publications (8). When physical data are digitized, valuable information and metadata risks being lost (17, 21).

Not everyone has access to the same data. Often data are proprietary, owned by pharmaceutical or industrial firms (1, 12, 2) or only searchable via expensive databases (19). Certain disciplines do not have domain repositories (5); multidisciplinary data repositories do not provide the search capabilities for specialized data, e.g. medical images (8). Existing data resources are also restricted by errors due to human indexing (13), unindexed information (7), or “ghost datasets” that are no longer curated or accessible (17). Research teams in numerous disciplines have built homegrown data repositories, but these resources are often not comprehensive (10, 13). Similarly, museum collections are built through networks and donations; curators decide what is included (17). Numerical data are often buried within larger numerical datasets (12) or the literature (19). Data are also not available for all regions of the world (17); if they are, the level of detail is not as good for certain regions (6).

## Evaluation

Evaluation is intertwined with data analysis and occurs throughout the data search process (13, 7, 6). Participants work extensively with data to determine its fitness for purpose (7). They also bring together information from multiple sources and perspectives to build fuller understandings and identify errors and biases (13, 2).

### *Social interactions in sense-making*

Participants seek out others to make sense of data, carefully choosing whom to contact. Some contact data authors directly (4, 5); others seek advice from colleagues (17, 7) or from carefully nurtured personal networks (2). Contacting experts is especially important when using data from outside of one’s discipline (14). One participant initiates collaborations in order to make sure he has team members with the necessary data expertise.

> I am used to working with experts from different areas of knowledge. For me it is usual to have partners with different expertise: biology, agronomy, economy…I know the language of LCA (life cycle assessment), not of electronics or agricultural biology. My limit is not the data that I cannot find, but people that can work with these data (16).

Data reuse or implementation in a new situation requires more than metadata and documentation (7, 6, 4). While metadata and documentation may provide enough information for a paper (7) or for background information (4), dialogues with data creators are imperative in ensuring appropriate and efficient reuse.

Participants also use a variety of information about the data in their evaluations (Table 6); these are combined with other strategies to build trust and establish data quality (Table 7).

| Discipline | Literature | Methodology/ collection conditions | Author | Source (repository, journal, etc.) | Documentation/ metadata | Data characteristics (well-structured, format) | Size of dataset | Time period |
|---|---|---|---|---|---|---|---|---|
| Acoustical engineering | 15 | 15 | 15 | | | | | |
| Business & finance | | | | 2 | | | | |
| Chemistry | | | | | | | | |
| Ecology - industrial & evolutionary | 4, 16 | 16 | 16 | 16 | 16 | 16 | | |
| Information science / Computer Science | 5 | 5, 8, 14, | 10, 14 | 7, 8, 10 | 5, 7, 8, 14 | 5, 8, 10 | 5, 8, 9, 10, 14 | 5, 10 |
| Librarians/ Literature managers | 20, 22 | | 22 | | 20 | | | |
| Medical statistics | 1 | 1 | | | | 1 | | |
| Molecular biology | 13 | | | 13 | | | | |
| Paleontology | | 17 | | | | 17 | | 17 |
| Popular culture | | | | | | | | 18 |
| Psychiatry, medicine, neuroscience | 12 | 12 | 12 | | | | | 12 |
| Psychology, Cognitive psychology, neuroscience | 11 | 11 | 11 | | 11 | 11 | 11 | 11 |
| Water resources | | | | 6 | | 6 | | 6 |
| Astronomy | | 20 | 20 | | | | | |

| Researchers | Support | Industry | Citizens |
|---|---|---|---|

*Table 6. Information used to evaluate data*

| | Trust | | | | | | Quality | | | | | | |
|---|---|---|---|---|---|---|---|---|---|---|---|---|---|
| Discipline | Author/ Annotator/ Indexer | Source (repository, journal, etc.) | Transparency | Social interactions | Methodology | Use | Lack of errors | Established through exploratory data analysis | Completeness | Source | Textual accuracy (e.g. lack of misspellings) | Methodology | Fitness for purpose |
| Acoustical engineering | | | | | | | 15 | | | 15 | | | |
| Business & finance | | 2 | | | | | | | | 2 | | | 2 |
| Chemistry | | | | | | | 19 | | | | | | |
| Ecology - industrial & evolutionary | | 16 | 16 | 4, 16 | 4 | 16 | | | | | 4 | | |
| Information science / Computer science | 7, 8, 14 | 10 | | 7, 8, 14 | | 5, 10 | 5, 20 | 7, 10, 14 | 5, 9 | | 10 | 14 | 7 |
| Librarians/ Literature managers | | | 20 | | | | | | | | | | |
| Medical statistics | | | | | 1 | | | | | | | 1 | |
| Molecular biology, medical devices | | | 13 | | | | 13 | | 13 | | | | |
| Paleontology | 17 | | 17 | 17 | | | 17 | | 17 | | | | |
| Popular culture | | | | | | | | | | | | | |
| Psychiatry, medicine, neuroscience | | | 12 | | | | | | | | | | |
| Psychology, Cognitive psychology, neuroscience | | | 11 | | 11 | | | 11 | | | | | |
| Water resources | | 6 | | | | | | 6 | 6 | | | | |
| Astronomy (reported) | 20 | | | | | | | | | | | | |

Researchers Support Industry Citizens

*Table 7. Information used to establish trust and data quality*

### *Information used to evaluate data*

The reputation, affiliation and size of repositories factor into reuse decisions. Some participants only use data from collaborators (7). Those in support roles rely heavily on the credibility of sources (20, 22); sometimes source information is the only evaluation criteria available to those outside of academia (2). The prolificacy (14), reputation (15, 22, 11), and expertise (8) of the data author are important for some participants, but not for others (1, 5).

Participants require details about data collection and handling. They need information regarding environmental conditions (17), geographic coordinates (6), instrumentation and calibration (15), creation parameters (15, 16, 14, 4, 8) and experimental protocols (4). Information about data provenance (14), processing (4) and statistical (1) or computational (20) tools is also used. Data must be well-structured, clearly labelled (11, 4), and in the desired format (10, 1) or resolution (6).

Participants find the needed information in metadata, documentation and codebooks, and the literature. Individuals in support roles rely exclusively on these sources to decide which information to pass on to end users. One participant finds information solely through complex literature searching, as she does not trust herself to evaluate medical data.

> I just would never trust myself enough; these are people's lives we are talking about. They (the customers) probably get more information than they need, but I think that it is helpful to them (22).

### *Trust*

We asked participants how they establish trust in secondary data. Some participants trust datasets commonly used in their disciplines or those used in peer-reviewed journals. Others need easy access to information about collection and analysis methods. Even small errors in the data or in the accompanying documentation will make researchers suspicious of the data (13). If a dataset is of sufficient size, worries about error are mitigated, as possible errors will be less significant. Errors can be identified by downloading and exploring data; if this process is facilitated, data are considered trustworthy (11).

The source of data, whether repository, journal, or governmental agency, is important in establishing trust. Social interactions with data owners provide a level of trust that cannot be easily replicated in an infrastructural system (17). Participants also consider the skill and reputation of the data annotator. This is used in conjunction with quality checks and knowledge of the data collection systems to develop trust (8).

Human and machine errors in indexing and information systems limit trustworthiness (13). Participants are aware that there is no such thing as a perfect dataset (13, 7), and that even seemingly trustworthy data may have been cleaned and presented in order to hide errors (4).

#### *Quality*

We asked how participants think about data quality. Most associate data quality with minimal error; strategies to evaluate potential error include checking methodologies (1), choosing data in peer-reviewed journals (15, 16), or finding large datasets that minimize the effects of possible errors (5, 10). Participants also engage in exploratory data analysis, such as performing basic count checks (14); qualitative statistical checks (11) and calculating ratios between variables (e.g. precipitation to stream flow) (6) to evaluate quality. Completeness of metadata fields and coverage also indicates quality, although a certain level of incompleteness is to be expected in some cases (13, 7, 6).

## Discussion

Having presented empirical evidence about data search practices, we now synthesize and discuss our findings through a conceptual framework positing data seeking as a contextual, socio-technical practice. Examining our findings using this perspective, we see that both the data seeker and data themselves are often narrowly conceptualized, particularly by system designers. We follow this discussion with a proposal for how the theoretical points we develop could be transformed into practical considerations for system developers.

### A Broader Understanding of the User

Both the literature study and interviews reveal that it is not enough to think of data users as researchers in a discipline with fixed practices. Communities, research interests, and practices are dynamic, at times influenced by the development of new research and analysis techniques (e.g. data science). Interdisciplinary projects create new communities and contexts, which necessitate new negotiations of data norms and blur disciplinary lines while enabling data discovery and reuse.

Not everyone who seeks data is a researcher. Librarians, literature managers, and students also seek data, using different strategies and evaluation methods. Individuals outside academia, including people working in industry and concerned citizens, are interested in finding and using data as well. The hint of another possible “user” is also emerging: the machine. As information retrieval systems develop to include proactive searching, some of the work currently done by humans may be automated in the future.

### A Broader Understanding of Data

Data needed for research are not always research data. Metadata, texts, server logs, device specifications, social media posts – all are used for foreground and background purposes in research but do not fall into what is traditionally thought of as “research data.” This finding reflects the idea that people define data differently,[46] perhaps as a result of how they intend to use them.

Applying our analytical perspective, it becomes possible to view data themselves as part of the dynamic ensemble of factors constituting context, with their very definition shaped by a user's intentions.

Data are also not always findable or reusable. Limitations in infrastructure, such as unstandardized metadata, un-curated datasets, or incomplete collections determine what data can be found, and thus reused. At the individual level, inconsistent data sharing practices, search abilities, social networks, and access rights hamper data search and reuse. Once data are found, there is no guarantee that users can access the resources (including humans) necessary to interpret and appropriately use the data.

While our findings support the idea of background and foreground data use (Wallis, 2013), we also find that data act as hubs for collaboration and creativity. Researchers form new collaborations in order to share, access and make sense of data. These collaborations help to grow professional networks and can inspire new approaches or future projects.

### Liminality

Data are not static, of interest only to the community where they are produced. They move between situations and communities, existing in different contexts and being adapted to different purposes. The pathways that data travel also depend heavily on context – the context of their creation and the contexts of discovery and reuse.

While data search exists as an independent practice, it is also liminal, at times situated in other practices and spanning their thresholds. When users are engaged in data analysis, for example, they also negotiate meaning. When they manage or share their own data, they discover new resources. When they engage with their professional networks, they both find and make sense of data.

### Data Search as a Socio-technical Practice

Some aspects of data seeking practices may seem clearly social, such as contacting authors or forming collaborations to access and understand data. Some aspects may seem clearly technical, such as retrieving data through an API or using exploratory data analysis.

The aim of socio-technical research is not to examine the social and the technical in isolation, but to examine the interactions that occur where the two intersect.[47] Applying this perspective to our findings, we see that data search practices are situated within and formed by interactions between the social and the technical spheres. For example, as users search for and evaluate data, they rely on metadata and documentation. Metadata schemas are created by humans; human practices and contexts also determine how those fields are populated. Users negotiate the social and technical worlds almost simultaneously, crossing the threshold between the two seamlessly, pointing again to the liminality of the process.

## Conclusion: Ramifications for System Design

Incorporating such theoretical understandings into the practical realm of system design can be valuable;[48] we therefore conclude by suggesting the following points for designers of data discovery systems to consider, before highlighting areas for future work.

There is great variation in data sharing and description practices. Designers could engage with disciplinary communities and repository managers to improve metadata standardization.

Systems could also incorporate techniques for enriching metadata automatically or consider how to operationalize best practices such as the FAIR data principles, as do Doorn and colleagues.[49]

Researchers find data in numerous repositories; discovery systems should therefore index both disciplinary and multidisciplinary repositories.

Data definitions and needs are changing. In order to support users' changing needs, systems could point to other data besides "research data." Disciplinary categories may also need rethinking to reflect the increasing interdisciplinarity of research.

Given the variety of users, needs, and search preferences, systems should support keyword searching, browsing, and include an API. Differentiated search interfaces for user groups and support for students or disciplinary novices could also be implemented. Interactive maps, or "macroscopes," providing visual overviews of repository contents[50,51] could provide this support.

Users discover data serendipitously – either through networks or when searching for other information. They also find data when engaging in data sharing or data management. Systems could be designed for serendipitous discovery and be integrated with infrastructures supporting other data practices.

Social interactions are used to locate, evaluate and develop trust in data. Data themselves can facilitate new collaborations. Designing ways to contact data authors and ranking datasets via social signals could support social interactions. More speculatively, integrating offline and online interactions around data, including links to in-person training opportunities, would be worth investigating.

The inherent social nature of data search exceeds what can be implemented in a discovery system and needs to be addressed by various stakeholders, including policy makers. Policies and guidelines are often drafted from perspectives that bury the complexity and cultural specificity of data sharing and reuse.[52,53] While guidelines such as the FAIR data principles recognize the importance of making data findable, accessible, interoperable, and reusable, the importance of embedded social communication and the relevance of data practices in domain-specific epistemic processes, particularly in reusability, still need to be made explicit and deserve further study.

Our principal contributions in this work bridge areas that are often disconnected. First, our work provides an example of how scientometric studies can inform and shape qualitative research. This connection between the quantitative and qualitative is further strengthened in our tabular presentation of our interview data. We also present evidence that data search is a complex phenomenon grounded in the interplay between technology and social practices, but not reducible to either. Finally, we connect the theoretical and practical realms, suggesting how our findings could be implemented in system design.

More remains to be done. In particular, there is a need to further connect social and technical research by integrating broad query log analyses with in-depth case studies. Applying existing models of information seeking behaviors to examine data seeking behaviors could also offer a way to explore similarities and differences in practices, perhaps leading to new models describing data search practices.

## Acknowledgements

KG developed the conceptual frameworks, collected and analyzed the data, and wrote the manuscript. HC, PG, AS, and SW contributed to theory development and editing.

## Funding

This work is part of the project *Re-SEARCH: Contextual Search for Research Data* and was funded by the NWO Grant 652.001.002.